\documentclass[twocolumn,english,aps]{revtex4}
\usepackage[T1]{fontenc}
\usepackage[latin1]{inputenc}

\makeatletter

\providecommand{\LyX}{L\kern-.1667em\lower.25em\hbox{Y}\kern-.125emX\@}

\makeatother
\begin{document}

\title{Magnetic effect from non-magnetic impurity in superconducting CuO\( _{2} \)
plane}

\author{V.M. Loktev}

\email{vloktev@gluk.org}

\affiliation{Bogolyubov Institute for Theoretical Physics, 14b Metrologichna str.,03143
Kiev-143, Ukraine}

\author{Yu.G. Pogorelov}

\email{ypogorel@fc.up.pt}

\affiliation{CFP/Departamento de Fisica, Unversidade do Porto, Rua do Campo Alegre
687, 4169-007 Porto, Portugal}

\begin{abstract}
We propose a new model for impurity center formed by a cathion substitute
for Cu, like Zn, in CuO\( _{2} \) planes. Its main effect on superconducting
electrons is due to the non-zero exchange field on O sites, neighbors
to the (non-magnetic) impurity. We discuss a strong suppression of
\( d \)-wave order parameter, a zero-energy resonance in local density
of states, and spin polarization of charge carriers, which can be
related to the experimentally observed effects in Zn-doped copper
oxides. These results are obtained \emph{without} using the unitary
scattering limit. 
\end{abstract}
\maketitle

It is recognized that there are two types of impurities in high-Tc
superconductors (HTSC): i) intrinsic or {}``own{}'', and ii) extrinsic
or {}``foreign{}''. The first type are the heterovalent impurities
or oxygen vacancies, that is the dopants. They supply charge carriers
into insulating antiferromagnetic (AFM) cuprate planes enabling their
metallization \cite{And1}, but also they act as scattering centers
for carriers. Previously, we showed \cite{LP1, LP2, LP3} that formation
of superconducting (SC) condensate, either of \( s \)- and \( d \)-
symmetry, is possible at low enough concentration \( c \) of such
impurities but is prevented by growing fluctuations of SC order parameter
at higher \( c \). 

The second type are the homovalent impurities, which only produce
scattering of existing charge carriers and so can depress the SC properties
of HTSC systems. They are well studied in common SC metals, where
it was stated by Anderson \cite{And} that non-magnetic impurities
have practically no effect on SC characteristics. At the same time,
even low concentration of paramagnetic ions can completely destroy
SC order, as initially shown in the Born approximation by Abrikosov
and Gor'kov \cite{AG} and then confirmed by many authors under more
general scope \cite{Maki, Shiba, Z-MH}.

But in the case of SC copper oxides, an apparent violation of this
so well theoretically based phenomenological principle was detected.
Thus, introducing of non-magnetic Zn\( ^{2+} \) ions instead of Cu\( ^{2+} \)
into the cuprate planes has a suppression effect on HTSC not weaker
but rather stronger than that by magnetic Ni\( ^{2+} \) ions \cite{Zhang, Bonn}.
This triggered an idea of viewing the non-magnetic impurity ions in
HTSC as extremely strong scatterers \cite{Chien} so that their perturbation
potential \( V_{imp} \) is the biggest energy parameter, treated
in the unitary limit: \( V_{imp}/W\gg 1 \) (\( W \) the bandwidth).
This concept was extensively elaborated \cite{Lee, Hirsch, Fehr},
the principal conclusions being the finite density of quasiparticle
states (DOS) at the Fermi level \( \varepsilon _{{\rm F}} \): \( \rho (\varepsilon \to \varepsilon _{{\rm F}})\to \rho _{u}\neq 0 \),
and the universal value of quasiparticle conductivity \( \sigma (\omega \to 0)\to \sigma _{u}\neq 0 \).
However, apart from the still existing controversies about those predictions
\cite{Atk}, it should be noted that, unlike the dopants, the foreign
impurity centers are formed in the CuO\( _{2} \) plane by a \emph{homovalent}
substitution (as Zn\( ^{2+} \) or Ni\( ^{2+} \) for Cu\( ^{2+} \)),
and it is problematic how they could produce such a strong perturbation
potential. Also we notice that the heterovalent non-magnetic scatterers
by dopants can not produce such effects \cite{LP2}.

This Letter is aimed on an alternative approach to the problem of
foreign impurities. It will be shown that irrespectively of the type
(magnetic or non-magnetic) of the cathion substitute in CuO\( _{2} \)
plane, the resulting center acts on charge carriers as \emph{magnetic}.
In accordance with the generally accepted notion, such center should
in fact strongly suppress SC order either of \( s \)- or \( d \)-type,
as was first qualitatively stated yet by Mahajan \emph{et al} \cite{Mah}.
We note that similar views on the effect of Zn impurities in HTSC
cuprates were expressed in some recent publications \cite{McFarl, Schr, Park},
though still focused on unitary scattering. Below we consider the
problem of isolated non-magnetic impurity ion in a CuO\( _{2} \)
plane and its local effects on the \( d \)-wave SC order parameter,
local density of states (LDOS), and the itinerant spin polarization.
Our treatment does not need using the unitary limit, nevertheless
the effects can be quite strong.

Fig. 1 shows a cathion impurity substitute for Cu in a CuO\( _{2} \)
plane, like real Zn, Fe, or Ni impurities in high-\( T_{c} \) compounds.
Associating the charge carriers mostly to O\( ^{-} \) holes, we conclude
that the main perturbation by such impurity (both magnetic and non-magnetic)
is due to the fact that its neighbor O sites occur in a non-zero exchange
field by Cu\( ^{2+} \) ions \cite{Wal, Jul}, which is equivalent
to the effect of magnetic impurity in a common superconductor. On
the other hand, there are no reasons to consider any sizeable spin-independent
perturbation from such isovalent impurity. The respective model Hamiltonian
is \( H=H_{sc}+H_{c}+H_{int} \). The unperturbed SC term \( H_{sc}=\sum _{\mathbf{k}}\psi _{\mathbf{k}}^{\dagger }(\xi _{\mathbf{k}}\hat{\tau }_{3}+\Delta _{\mathbf{k}}\hat{\tau }_{1})\psi _{\mathbf{k}} \)
couples the Nambu spinors \( \psi _{\mathbf{k}}^{\dagger }=(a_{\mathbf{k},\uparrow }^{\dagger },a_{-\mathbf{k},\downarrow }) \)
with Pauli matrices \( \hat{\tau }_{i} \). The normal dispersion
\( \xi _{\mathbf{k}}=W(2-\cos ak_{x}-\cos ak_{y})/4-\varepsilon _{{\rm F}} \)
leads to the density of states (DOS) \( \rho _{0}=4/(\pi W) \). The
\( d \)-wave gap function \( \Delta _{\mathbf{k}}=\Delta \theta (\varepsilon _{{\rm D}}^{2}-\xi _{\mathbf{k}}^{2})\gamma _{\mathbf{k}}/\gamma _{m} \)
includes the BCS-shell factor \( \theta (\varepsilon _{{\rm D}}^{2}-\xi _{\mathbf{k}}^{2}) \)
with the {}``Debye energy{}'' \( \varepsilon _{{\rm D}} \), the
symmetry factor \( \gamma _{\mathbf{k}}=\cos ak_{x}-\cos ak_{y} \)
with maximum absolute value \( \gamma _{m}=\pi \varepsilon _{{\rm F}}\rho _{0} \),
and the gap parameter\begin{equation}
\label{delta}
\Delta =VN^{-1}\sum _{\mathbf{k}}\gamma _{\mathbf{k}}\langle a_{-\mathbf{k},\downarrow }a_{\mathbf{k},\uparrow }\rangle 
\end{equation}
 where \( V \) is the attraction between two carriers with opposite
spins on neighbor O sites. The term \( H_{c}=-hS_{z} \) models the
(AFM) correlation between the impurity center and its environment,
where \( h\sim J_{dd} \), the Cu-Cu exchange constant, and \( \mathbf{S} \)
is the spin of a fictitious {}``magnetic impurity{}''. It can be
seen as a cluster of four 1/2 spins of Cu nearest neighbors to real
non-magnetic impurity (Fig. 1). In reality, its quantization axis
\( z \) is only defined over time periods no longer than \( \tau _{s}\sim \hbar \xi _{s}/(aJ_{dd})\sim 10^{-13} \)
s for spin correlation length \( \xi _{s}\sim a/\sqrt{x} \) \cite{Birg}
and doping levels \( x\sim 0.1 \) (this also agrees with the NMR
data \cite{McFarl}). However this \( \tau _{s} \) is much longer
than typical electronic times \( \sim \hbar /\varepsilon _{{\rm F}}\sim 10^{-15} \)
s for HTSC compounds. For \( h>0 \) we have \( \langle S_{z}\rangle \equiv s \)
and \( 0<s<S \), which accounts for the short-range AFM order, whereas
\( s\to 0 \) in the paramagnetic limit \( \beta h\ll 1 \).

We separate the spin-dependent interaction between charge carriers
and impurity into three parts:\begin{equation}
\label{int}
H_{int}=H_{int}^{{\rm MF}}+H_{int}^{\parallel }+H_{int}^{\perp },
\end{equation}
where \[
H_{int}^{{\rm MF}}=JsN^{-1}\sum _{\mathbf{k},\mathbf{k}'}\sum _{\sigma =\pm }\alpha _{j,\mathbf{k}}\alpha _{j,\mathbf{k}'}\sigma a_{\mathbf{k}',\sigma }^{\dagger }a_{-\mathbf{k},\sigma }\]
 is the {}``mean-field{}'' (MF) polarization of carrier spins by
the impurity center, and \[
H_{int}^{\parallel }=JN^{-1}\sum _{\mathbf{k},\mathbf{k}'}\sum _{\sigma =\pm }\alpha _{j,\mathbf{k}}\alpha _{j,\mathbf{k}'}\sigma (S_{z}-s)a_{\mathbf{k}',\sigma }^{\dagger }a_{\mathbf{k},\sigma },\]
 \[
H_{int}^{{\rm MF}}=JN^{-1}\sum _{\mathbf{k},\mathbf{k}'}\sum _{\sigma =\pm }\alpha _{j,\mathbf{k}}\alpha _{j,\mathbf{k}'}S_{\sigma }a_{\mathbf{k}',-\sigma }^{\dagger }a_{\mathbf{k},\sigma },\]
 are their interactions with longitudinal and transversal fluctuations
of \( \mathbf{S} \). In the paramagnetic limit: \( s\to 0 \), Eq.
\ref{int} is reduced to the common Kondo interaction \cite{Kon, Nag}.
For definiteness, the Cu-O \( p \)-\( d \) exchange parameter \( J \)
is considered positive. The functions\begin{eqnarray*}
\alpha _{1,\mathbf{k}}=2\cos \frac{ak_{x}}{2}\cos \frac{ak_{y}}{2},\quad \alpha _{2,\mathbf{k}}=2\cos \frac{ak_{x}}{2}\sin \frac{ak_{y}}{2}, &  & \\
\alpha _{3,\mathbf{k}}=2\sin \frac{ak_{x}}{2}\cos \frac{ak_{y}}{2},\quad \alpha _{4,\mathbf{k}}=2\sin \frac{ak_{x}}{2}\sin \frac{ak_{y}}{2}, &  & 
\end{eqnarray*}
 realize 1D irreducible representations of the \( C_{4v} \) point
group (of the plaquette surrounding the impurity), so that \( N^{-1}\sum _{\mathbf{k}}\alpha _{j,\mathbf{k}}\alpha _{j',\mathbf{k}}=\delta _{jj'} \).
Distinctive features of the perturbation, Eq. \ref{int}, compared
to the commonly used impurity models, are: i) its spatial extension
expressed by the factors \( \alpha _{j,\mathbf{k}} \), ii) additional
degrees of freedom by spin \( \mathbf{S} \), and iii) coupling of
\( \mathbf{S} \) to the local AFM correlations.

In principle, this impurity center can produce yet another perturbation,
due to a possible role of AFM correlated Cu\( ^{2+} \) spins in the
SC coupling between charge carriers. Lacking one such spin would locally
perturb the \( \Delta _{\mathbf{k}}\hat{\tau }_{1} \) term in \( H_{sc} \)
by some expansions in \( \alpha _{j,\mathbf{k}}\alpha _{j,\mathbf{k}'} \).
This can influence the SC order, alike the simpler case of point-like
perturbation of \( s \)-wave SC coupling \cite{Pog}. However, for
simplicity, we leave this kind of perturbation for a separate study.

We calculate the averages, like Eq. \ref{delta}, by simple spectral
formula at \( T=0 \):\begin{equation}
\label{av}
\langle ab\rangle =\pi ^{-1}\int _{0}^{\varepsilon _{{\rm F}}}\Im \langle \langle b|a\rangle \rangle _{\varepsilon }d\varepsilon ,
\end{equation}
where \( \Im f(\varepsilon )=\lim _{\delta \to 0}[f(\varepsilon -i\delta )-f(\varepsilon +i\delta )]/2 \)
and \( \langle \langle b|a\rangle \rangle _{\varepsilon \pm i\delta } \)
are the retarded and advanced two-time Green functions (GF's). The
relevant GF matrix \( \hat{G}_{\mathbf{k},\mathbf{k}'}=\langle \langle \psi _{\mathbf{k}}|\psi _{\mathbf{k}'}^{\dagger }\rangle \rangle  \)
in absence of impurity perturbation (\( J=0 \)) is momentum-diagonal:
\( \hat{G}_{\mathbf{k},\mathbf{k}'}=\delta _{\mathbf{k},\mathbf{k}'}\hat{G}_{\mathbf{k}} \),
with \( \hat{G}_{\mathbf{k}}=(\varepsilon -\xi _{\mathbf{k}}\hat{\tau }_{3}-\Delta _{\mathbf{k}}\hat{\tau }_{1})^{-1} \).
The same expression holds for the momentum-diagonal GF \( \hat{G}_{\mathbf{k},\mathbf{k}} \)
in presence of single impurity, whose effect \( \sim 1/N \) is negligible
for this quantity. However it is only this small impurity effect that
gives rise to a momentum-non-diagonal GF's \( \hat{G}_{\mathbf{k},\mathbf{k}'} \).
They are found from the equation of motion \( \hat{G}_{\mathbf{k},\mathbf{k}'}=JN^{-1}\sum _{\mathbf{k}'',j}\alpha _{j,\mathbf{k}}\hat{G}_{\mathbf{k}}(s\hat{G}_{\mathbf{k}'',\mathbf{k}'}+\hat{G}_{\mathbf{k},\mathbf{k}'}^{(z)}+\hat{G}_{\mathbf{k},\mathbf{k}'}^{(-)})\alpha _{j,\mathbf{k}''} \)
including three scattered GF's: the MF one \( \hat{G}_{\mathbf{k}'',\mathbf{k}'} \),
the longitudinal \( \hat{G}_{\mathbf{k}'',\mathbf{k}'}^{(z)}=\langle \langle \psi _{\mathbf{k}''}(S_{z}-s)|\psi _{\mathbf{k}'}^{\dagger }\rangle \rangle  \)
and the transversal \( \hat{G}_{\mathbf{k}'',\mathbf{k}'}^{(-)}=\langle \langle \bar{\psi }_{\mathbf{k}''}S_{-})|\psi _{\mathbf{k}'}^{\dagger }\rangle \rangle  \)
with \( \bar{\psi }_{\mathbf{k}}^{\dagger }=(a_{\mathbf{k},\downarrow }^{\dagger },a_{-\mathbf{k},\uparrow }) \).
The two last terms are analogous to the well known Nagaoka's \( \Gamma  \)-term
\cite{Nag, Z-MH} and treating them with a similar decoupling procedure
gives:\begin{eqnarray*}
\hat{G}_{\mathbf{k},\mathbf{k}'}^{(z)}=\frac{J\Sigma ^{2}}{N}\sum _{\mathbf{k}'',j}\alpha _{j,\mathbf{k}}\hat{G}_{\mathbf{k}}\hat{G}_{\mathbf{k}'',\mathbf{k}'}\alpha _{j,\mathbf{k}''},\quad \quad \quad \quad \quad  &  & \\
\hat{G}_{\mathbf{k},\mathbf{k}'}^{(-)}=\frac{J}{N}\sum _{\mathbf{k}'',j}\alpha _{j,\mathbf{k}}\hat{G}_{\mathbf{k}}(\varepsilon +h)\hat{X}_{\mathbf{k}''}\hat{G}_{\mathbf{k}'',\mathbf{k}'}\alpha _{j,\mathbf{k}''}, &  & 
\end{eqnarray*}
where \( \Sigma ^{2}=\langle S_{z}^{2}\rangle -s^{2} \), \( \hat{X}_{\mathbf{k}}=S(S+1)-s(s+1)-\Sigma ^{2}+(1+2\xi _{\mathbf{k}}/E_{\mathbf{k}})\hat{\tau }_{3} \),
\( E_{\mathbf{k}}=\sqrt{\xi _{\mathbf{k}}^{2}+\Delta _{\mathbf{k}}^{2}} \),
and one energy argument is shifted: \( \varepsilon \to \varepsilon +h \),
due to the AFM stiffness. Finally, we obtain the decoupled equation
of motion:\[
\hat{G}_{\mathbf{k},\mathbf{k}'}=N^{-1}\sum _{\mathbf{k}'',j}\alpha _{j,\mathbf{k}}\hat{G}_{\mathbf{k}}[Js+J^{2}(\Sigma ^{2}\hat{G}_{j}+\hat{X}_{j})]\hat{G}_{\mathbf{k}'',\mathbf{k}'}\alpha _{j,\mathbf{k}''}\]
 where \( \hat{G}_{j}=N^{-1}\sum _{\mathbf{k}}\alpha _{j,\mathbf{k}}^{2}\hat{G}_{\mathbf{k}} \),
\( \hat{X}_{j}=N^{-1}\sum _{\mathbf{k}}\alpha _{j,\mathbf{k}}^{2}\hat{G}_{\mathbf{k}}(\varepsilon +h)\hat{X}_{\mathbf{k}} \),
and its standard iteration yields in the result:\begin{equation}
\label{Tmat}
\hat{G}_{\mathbf{k},\mathbf{k}'}=N^{-1}\sum _{j}\alpha _{j,\mathbf{k}}\hat{G}_{\mathbf{k}}\hat{T}_{j}\hat{G}_{\mathbf{k}'}\alpha _{j,\mathbf{k}'},
\end{equation}
with the partial T-matrices \( \hat{T}_{j}=[Js+J^{2}(\Sigma ^{2}\hat{G}_{j}+\hat{X}_{j})][1-Js-J^{2}(\Sigma ^{2}\hat{G}_{j}+\hat{X}_{j})]^{-1} \).
By the definition of our model, the parameter \( Js \) is positive.
It is interesting to trace the behavior of \( \hat{T}_{j} \) in the
two characteristic limits for AFM correlations between Cu\( ^{2+} \)spins.

In the paramagnetic limit: \( h\to 0 \), \( s\to 0 \), we have \( \Sigma ^{2}\to S(S+1)/3 \)
and \( \hat{X}_{j}\to 2S(S+1)/3-N^{-1}\sum _{\mathbf{k}}\alpha _{j,\mathbf{k}}^{2}(1+2\xi _{\mathbf{k}}/E_{\mathbf{k}})\hat{G}_{\mathbf{k}}\hat{\tau }_{3} \).
In neglect of the small last term we arrive at:\[
\hat{T}_{j}\to J^{2}S(S+1)\hat{G}_{j}[1-J^{2}S(S+1)\hat{G}_{j}]^{-1}\]
generalizing the known results \cite{AG, Z-MH} for the case of extended
impurity center.

Another limit, fully polarized, \( h\to \infty  \), \( s\to S \),
corresponds to \( \Sigma ^{2}\to 0 \), \( \hat{X}_{j}\to 0 \) and
results in\begin{equation}
\label{pol}
\hat{T}_{j}\to JS(1-JS\hat{G}_{j})^{-1}
\end{equation}
which is only due to the effect of MF magnetic scattering. The obvious
validity condition for this limit, \( JS\gg k_{{\rm B}}T \), well
applies in the SC phase at \( T<T_{c}\sim \Delta /k_{{\rm B}} \),
so we use Eq. \ref{pol} for the T-matrices in what follows.

The local SC correlation is characterized by the average \( \Delta _{12}=2V\langle a_{\delta _{1},\downarrow }a_{\delta _{2},\uparrow }\rangle  \)
(see Fig. 1) where a site operator \( a_{\mathbf{n},\sigma } \) is
expressed through band operators: \( a_{\mathbf{n},\sigma }=N^{-1/2}\sum _{\mathbf{k}}{\rm e}^{i\mathbf{k}\cdot \mathbf{n}}a_{\mathbf{k},\sigma } \).
Since the phase of \( \Delta _{\mathbf{k}} \) is chosen zero, \( \Delta _{12} \)
is real. For \( J=0 \), this average does not differ from the uniform
gap parameter:\[
\Delta _{12}\to \frac{2V}{N}\sum _{\mathbf{k}}\langle a_{-\mathbf{k},\downarrow }a_{\mathbf{k},\uparrow }\rangle {\rm e}^{i\mathbf{k}\cdot (\delta _{2}-\delta _{1})}=\Delta ,\]
whereas for \( J\neq 0 \) the maximum perturbation of SC order near
the impurity is given by the suppression parameter \( \eta _{sup}=1-\Delta _{12}/\Delta  \).
Its value is confined between 0 (for pure SC) and 1 (for complete
local suppression of SC order), and it only results from non-diagonal
GF's:\begin{eqnarray}
\eta _{sup}=\frac{2V}{N\Delta }\sum _{\mathbf{k},\mathbf{k}'\neq \mathbf{k}}\langle a_{-\mathbf{k},\downarrow }a_{\mathbf{k}',\uparrow }\rangle {\rm e}^{i(\mathbf{k}\cdot \delta _{2}-\mathbf{k}'\cdot \delta _{1})}= &  & \label{hsupT} \\
=\frac{V}{2\pi \Delta }\sum _{j}(-1)^{j}\int _{-\infty }^{0}d\varepsilon {\rm Im\, Tr}\hat{G}_{j}\hat{T}_{j}\hat{G}_{j}\hat{\tau }_{1}, &  & \nonumber 
\end{eqnarray}
where the trace is in Nambu indices and Eqs. \ref{av},\ref{Tmat}
were used. Expansion of \( \hat{G}_{j} \) in Pauli matrices only
contains \( \hat{\tau }_{1} \)terms at \( j=2,3 \) (by the parity
of \( \alpha _{j,\mathbf{k}} \) with respect to the permutation \( k_{x}\leftrightarrow k_{y} \)):
\( JS\hat{G}_{2,3}=A+B\hat{\tau }_{3}\pm C\hat{\tau }_{1} \). Hence
only \( j=2,3 \) actually contribute in Eq. \ref{hsupT} by\begin{equation}
\label{hsup}
\eta _{sup}=\frac{V\varepsilon _{{\rm F}}\rho _{0}^{2}}{4}\int _{0}^{\varepsilon _{{\rm F}}}\frac{F(\varepsilon )}{\varepsilon }d\varepsilon ,
\end{equation}
where

\[
F(\varepsilon )=\frac{16\varepsilon }{\pi JS\varepsilon _{{\rm F}}\Delta \rho _{0}^{2}}{\rm Im}[1-\frac{1}{(1-A)^{2}-B^{2}-C^{2}}]C,\]
and the complex coefficients \( A,B,C \) as functions of energy are
estimated in the relevant range \( \Delta <\varepsilon <\varepsilon _{{\rm F}} \),
setting \( E_{\mathbf{k}}\approx \xi _{\mathbf{k}}\approx Wa(k^{2}-k_{{\rm F}}^{2})/8 \):\begin{eqnarray*}
A\approx \frac{JS\rho _{0}\varepsilon }{8}(\ln |\frac{\varepsilon _{{\rm F}}+\varepsilon }{\varepsilon _{{\rm F}}-\varepsilon }|+2i\pi ),\quad \quad \quad \quad  &  & \\
B\approx \frac{JS\rho _{0}}{8}\ln \frac{\varepsilon _{{\rm F}}^{2}-\varepsilon ^{2}}{(W-\varepsilon _{{\rm F}})^{2}},\quad \quad \quad  &  & \\
C\approx \frac{\pi JS\rho _{0}\varepsilon _{{\rm F}}\Delta }{E\varepsilon }[\ln |\frac{\varepsilon _{{\rm D}}+\varepsilon }{\varepsilon _{{\rm D}}-\varepsilon }|+i\pi \theta (\varepsilon _{{\rm D}}-\varepsilon )]. &  & 
\end{eqnarray*}
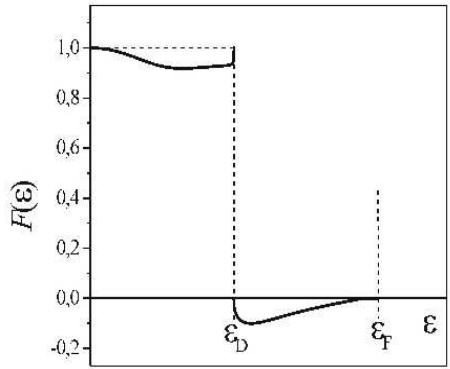
Numeric analysis of these expressions with realistic parameter values:
\( W\sim 2 \) eV, \( JS\rho _{0}\sim 1 \), \( \varepsilon _{{\rm F}}\sim 0.3 \)
eV, \( \varepsilon _{{\rm D}}\sim 0.15 \) eV, shows that the main
contribution into the integral, Eq. \ref{hsup}, comes from the BCS
shell \( \Delta <\varepsilon <\varepsilon _{{\rm D}} \), where we
have: \( F(\varepsilon )\approx 1,\: 0<1-F(\varepsilon )\ll 1 \),
while \( 0<-F(\varepsilon )\ll 1 \) out of this shell, \( \varepsilon _{{\rm D}}<\varepsilon <\varepsilon _{{\rm F}} \)
(Fig. 2). Taking in mind the relation \( \ln (\varepsilon _{{\rm D}}/\Delta )\approx W/(V\gamma _{m})=W^{2}/(4V\varepsilon _{{\rm F}}) \)
which follows from Eq. \ref{delta}, we obtain from Eq. \ref{hsup}
\emph{almost complete} suppression: \( \eta _{sup}\approx 1 \). A
small residual part \( 0<1-\eta _{sup}\ll 1 \) is only due to a small
negative deviation of \( F(\varepsilon ) \) from unity within the
shell and to a small negative out-of-shell contribution. Thus, for
the above indicated choice of parameters we have \( \eta _{sup}\approx 96\% \).
The value of \( 1-\eta _{sup} \) yet diminishes with growing \( JS \),
but it should be stressed that no unitary limit \( JS\rho _{0}\gg 1 \)
is needed to get such a strong effect. 

The decay of this maximum effect with separation \( \mathbf{R} \)
from the impurity is given, in similarity with Eq. \ref{hsupT}, by\begin{equation}
\label{etaR}
\eta _{sup}(\mathbf{R})=\frac{V}{\pi \Delta }\int _{-\infty }^{0}d\varepsilon {\rm Im\, Tr}\hat{G}_{2}(\mathbf{R})\hat{T}_{2}\hat{G}_{2}(\mathbf{R})\hat{\tau }_{1},
\end{equation}
 Here the matrix \( \hat{G}_{j}(\mathbf{R})=N^{-1}\sum _{\mathbf{k}}{\rm e}^{i\mathbf{k}\cdot \mathbf{R}}\alpha _{j,\mathbf{k}}\hat{G}_{\mathbf{k}} \),
and for \( R\gg \xi _{\pm }=a\sqrt{W/[8(\varepsilon _{{\rm F}}\pm \varepsilon )]} \)
it is mainly contributed by two saddle points in the complex \( k \)-plane:
\( \pm \xi _{\pm }^{-1}-iR^{-1} \), hence all its matrix elements
decay asymptotically like \( \cos (R/\xi _{\pm })/\sqrt{R/\xi _{\pm }} \),
and the non-diagonal elements contain yet the anisotropic factor \( (\Delta /\varepsilon )\cos 2\psi  \)
where \( \psi =\arctan R_{y}/R_{x} \). However, for the energies
\( \varepsilon \sim \varepsilon _{{\rm D}} \) relevant here, such
anisotropy is less pronounced than that in the limit \( \varepsilon \to 0 \)
considered by Balatsky \emph{et al} \cite{Bal}. Integration in Eq.
\ref{etaR} results in asymptotic \( \eta _{sup}(\mathbf{R})\approx \eta _{sup}\sqrt{W/(8\varepsilon _{{\rm F}})}\sum _{i=\pm }(u_{i}+f_{i}\cos 2\psi +h_{i}\cos 4\psi )(a/R) \)
where \( h_{i}\ll f_{i}\sim u_{i}\sim 1 \). This angular dependence
resembles that for LDOS around Zn impurity, suggested by Haas and
Maki from continuous Bogolyubov-de Gennes equations \cite{Haas},
while the anomalously slow radial decay should enhance the overall
suppression of SC order. 

Besides the considered local suppression of the order parameter, related
to the non-diagonal (in Nambu indices) elements of GF's \( \hat{G}_{\mathbf{k},\mathbf{k}'} \),
there are also local effects related to their diagonal elements. Thus,
the variation of local DOS (LDOS): \( \rho (\mathbf{n})=(\pi N)^{-1}\sum _{\mathbf{k},\mathbf{k}'\neq \mathbf{k}}{\rm Im\, Tr\, e}^{i(\mathbf{k}-\mathbf{k}')\cdot \mathbf{n}}\hat{G}_{\mathbf{k},\mathbf{k}'}\hat{\tau }_{3} \),
attains its maximum at \( \mathbf{n}={\bf \delta } \), nearest neighbor
sites to the impurity, and Eq. \ref{Tmat} mainly contributes there
by \( j=1 \): \( \rho ({\bf \delta })\approx {\rm Im\, Tr}\hat{G}_{1}({\bf \delta })\hat{T}_{1}\hat{G}_{1}({\bf \delta })\hat{\tau }_{3} \).
The relevant GF's are \( \hat{G}_{1}({\bf \delta })\approx 2N^{-1}\sum _{\mathbf{k}}\hat{G}_{\mathbf{k}}=2(g_{0}+g_{3}\hat{\tau }_{3}) \),
where \( g_{0}\approx \rho _{0}\varepsilon [\varepsilon _{{\rm F}}^{-1}+i(\pi /\Delta )\arcsin (\Delta /\varepsilon )] \)
and \( g_{3}\approx \rho _{0}\ln (W/\varepsilon _{{\rm F}}) \) \cite{LP2}.
Considered as a function of energy, \( \rho ({\bf \delta }) \) displays
a very sharp resonance in the denominator of \( \hat{T}_{1} \): \( {\rm Re}[1-2JSg_{0}(\varepsilon )]^{2}-(2JSg_{3})^{2}\to 0 \)
at \( \varepsilon \to 0 \) (cf. Fig. 3) with the observed peak in
the related tunnel conductivity \cite{Pan}) if the impurity perturbation
parameter \( J \) is close to \( J_{cr}=1/(2Sg_{3}) \). This refers
to a \emph{fine tuned} rather than unitary perturbation \( J \) and
agrees with its choice made to estimate \( \eta _{sup} \). The \( j=1 \)
contribution also dominates in the Kondo-like local polarization of
itinerant spins: \( m({\bf \delta })\approx \int _{0}^{\varepsilon _{{\rm F}}}d\varepsilon {\rm Im\, Tr\, e}^{i(\mathbf{k}-\mathbf{k}')\cdot \mathbf{n}}\hat{G}_{1}({\bf \delta })\hat{T}_{1}\hat{G}_{1}({\bf \delta }) \),
which should explain the observed enhancement of exchange fields on
\( ^{63} \)Cu \cite{Wal} and \( ^{89} \)Y \cite{Mah} nuclei close
to Zn impurities. A more detailed treatment of these phenomena will
be presented elsewhere.

The proposed model can be equally applied to isovalent substitutes
for Cu, which are magnetic themselves, as Ni\( ^{2+} \) or Fe\( ^{2+} \).
But, since the net MF on neighbor O sites in this case is due to incomplete
AFM compensation of exchange fields by different magnetic ions, the
perturbation parameter \( JS \) may be \emph{weaker} than that for
non-magnetic Zn\( ^{2+} \) and so the resulting suppression of SC
order, as is observed in the experiment \cite{Bonn}.

In conclusion, we developed a microscopic model of spin dependent
perturbation on charge carriers in CuO\( _{2} \) planes, produced
by a non-magnetic substitute for Cu. An almost complete suppression
of \( d \)-wave order parameter at nearest neighbor sites to the
impurity atom is obtained, as a result of parallel alignment of carrier
spins in the exchange field \( JS \) by non-compensated Cu\( ^{2+} \)
spins, and this strong effect is achieved with moderate \( JS \)
values. It decays with distance from impurity rather slowly, which
can explain the fast destroying of SC order in cuprates already at
low Zn concentration. The model also provides explanation for other
local effects, such as a sharp resonance of LDOS and local polarization
of charge carrier spins close to impurity.

\begin{acknowledgments}
We acknowledge the partial support for this work from Portuguese Fundação
de Ciência e Tecnologia through the project 2/2.1/FIS/302/94 and from
Swiss Science Foundation under SCOPES-project 7UKPJ062150.00/1.
\end{acknowledgments}

\end{document}